\documentclass[letterpaper, 10 pt, conference]{ieeeconf}  

\IEEEoverridecommandlockouts                              

\usepackage{graphics} 
\usepackage{epsfig} 
\usepackage{mathptmx} 
\usepackage{times}

\usepackage{amsthm}
\usepackage{amsmath} 
\usepackage{amssymb}  

\title{\LARGE \bf
A safety governor for learning explicit MPC controllers from data}

\author{Anjie Mao$^{1}$, Zheming Wang$^{1}$, Hao Gu$^{2}$, Bo Chen$^{1}$, and Li Yu$^{1}$
\thanks{This work is supported in part by the National Natural Science Foundation of China under Grant 62303416 and W2421028.}
\thanks{$^{1}$Anjie Mao, Zheming Wang, Bo Chen and Li Yu are with the Department of Automation, Zhejiang
 University of Technology, Hangzhou 310023, China (e-mail: maoanjie@aliyun.com;
wangzheming@zjut.edu.cn; bchen@
 aliyun.com; lyu@zjut.edu.cn). }%
 \thanks{$^{2}$ Hao Gu is
with Zhejiang Guoli Security Technology Co., Ltd (e-mail: guhao@glsec.com.cn). }%
}

\usepackage{amsmath,amssymb,amsthm,mathtools}
\usepackage{algorithm,algorithmic}
\usepackage{xcolor}
\newtheorem{theorem}{Theorem}
\newtheorem*{theorem*}{Theorem}

\newtheorem{proposition}{Proposition}
\newtheorem*{proposition*}{Proposition}

\newtheorem{definition}{Definition}
\newtheorem*{definition*}{Definition}

\newtheorem*{assumption*}{Assumption}

\newtheorem{remark}{Remark}
\newtheorem*{remark*}{Remark}

\begin{document}

\maketitle
\thispagestyle{empty}
\pagestyle{empty}

\begin{abstract}

We tackle neural networks (NNs) to approximate model predictive control (MPC) laws. We propose a novel learning-based explicit MPC structure, which is reformulated into a dual-mode scheme over maximal constrained feasible set. The scheme ensuring the learning-based explicit MPC reduces to linear feedback control while entering the neighborhood of origin. We construct a safety governor to ensure that learning-based explicit MPC satisfies all the state and input constraints. Compare to the existing approach, our approach is computationally easier to implement even in high-dimensional system. The proof of recursive feasibility for the safety governor is given. Our approach is demonstrated on numerical examples.

\end{abstract}

\section{INTRODUCTION}

Model predictive control (MPC) is an advanced control strategy that has been extensively utilized in industrial processes \cite{qin2003survey}. More recently, MPC have become popular to solve robot control problems owing to its capability in dealing with system constraints and optimizing control performance. MPC operates by repeatedly solving an optimization problem, leveraging updated state information in real-time. Despite the advantages of MPC, the shift from process industry applications to robotics introduces formidable challenges, primarily due to the drastic reduction in available computation time, transitioning from hours to mere milliseconds \cite{chen2022large}.

 In recent decades, substantial efforts have been directed towards alleviating the computational burden of MPC. Existing approaches can be broadly categorized into two main groups. The first involves developing bespoke, numerically efficient optimization solvers that capitalize on the inherent structure of the MPC problem, e.g., \cite{jerez2014embedded,patrinos2013accelerated}. Another major and promising approach is explicit model predictive control (EMPC), involves computing the optimal control policy offline as a piecewise-affine function of the state variables across different regions through parametric optimization, e.g., \cite{bemporad2002explicit,bemporad2002model}. EMPC offers several key advantages, including significantly faster online computations, precise bounds on worst-case execution time, and the generation of simple, verifiable control policies. These benefits make EMPC particularly well-suited for embedded applications where real-time performance and reliability are critical.  Nevertheless, in linear systems, the memory allocation and computational complexity associated with storing and evaluating the state
look-up table exhibit exponential growth. This escalation becomes particularly prohibitive when extending the prediction horizon or increasing the number of system constraints \cite{kvasnica2011clipping}. Moreover, extending this approach to nonlinear systems is not straightforward.

 One approach to mitigate the computational challenges associated with EMPC is to develop an approximate controller. In \cite{kvasnica2011clipping}, MPC laws are approximated via linear combination of basis functions. Meanwhile, \cite{jones2010polytopic} employs a double-description method to construct piecewise-affine approximations of the value function. In recent years, the widespread adoption of neural networks (NNs) has been propelled by their exceptional function approximation capabilities. Theoretical studies have established that NNs, when equipped with adequate network depth and width, possess the universal approximation property, enabling them to approximate any continuous nonlinear function with arbitrary precision \cite{hornik1990universal}. Since the concept of approximating MPC laws using NNs with a single hidden layer was introduced by Parisini et al. \cite{parisini1995receding}, numerous efforts have been made to explore learning-based approximations of MPC.  Notable contributions by 
 Chen et al. \cite{chen2018approximating},
 Karg et al. \cite{karg2020efficient}, 
 Montufar et al. \cite{montufar2014number},
Hertneck et al. \cite{hertneck2018learning}, 
 and Lucia et al. \cite{lucia2018deep}.

In most approximation scenarios, the MPC problems are initially solved offline. Subsequently, the resulting data pairs, which comprise initial states and their corresponding optimal control inputs, are utilized to approximate the MPC laws. However, a significant limitation of this approach lies in the approximated controller's propensity to violate constraints and accumulate approximation errors.
A widely adopted strategy to ensure the safety of the approximated controller involves projecting its control inputs onto a secure region, such as the maximal control invariant (MCI) set and feasibility origin of MPC \cite{chen2018approximating}. However, the determination of such sets is an NP-hard complete problem in high dimension \cite{2008On}. Current approximation methods for such sets, such as constructing the convex hull of feasible MPC solutions as control-invariant sets \cite{karg2020efficient}, may introduce approximation errors that compromise the theoretical feasibility guarantees of the original approach \cite{chen2018approximating}.

\begin{figure*}[!htb]
  \centering
  \includegraphics[width=0.98\linewidth]{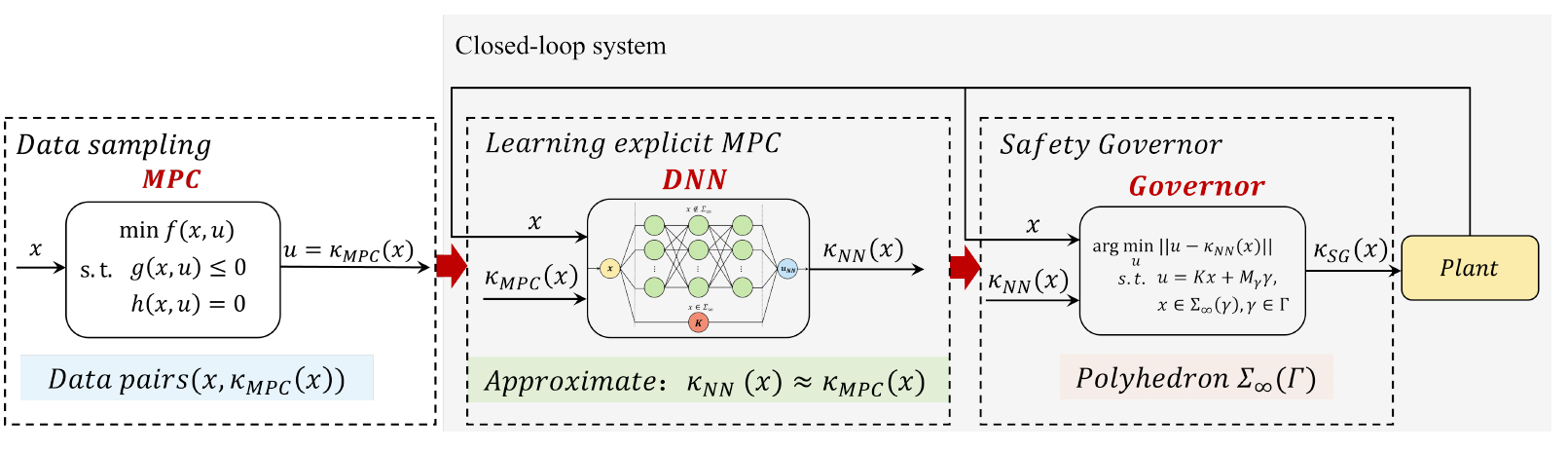}
  \caption{Diagram of the proposed framework}
  \label{fig: Diagram of the proposed framework}
\end{figure*} 

    In this paper, we discuss how to learn MPC law from data while guaranteeing safety. A sketch of the main ideas is given in Figure \ref{fig: Diagram of the proposed framework}. First, we design a standard MPC for a constrained discrete-time linear time-invariant system. The resulting MPC law is sampled offline for suitable states and learned (approximated) via a designed learning-based explicit MPC, in which a novel network structure with dual-mode scheme is proposed. A safety governor is constructed to ensure that learning-based explicit MPC satisfies all the state and input constraints. Compare to the approach in \cite{chen2018approximating}, our feasible origin is much easier to compute even in high-dimensional system.
    Numerical examples of low-order and high-dimensional systems demonstrate the control performance of our approach.

 \emph{Contributions}: (1) A novel learning-based explicit MPC structure is constructed, which has a novel dual-mode scheme can erase the approximation errors nearby the system origin. (2) An approach named safety governor is proposed, which guarantee the safety of the trained learning-based explicit MPC even with great approximation errors.

 \emph{Notation}: $\mathbb{R}$ and $\mathbb{Z}^+$ denote the real and non-negative integer numbers, respectively. $\mathbb{R}^n$ denotes the set of $n$-dimensional real vectors.
 The superscript $\top$ denotes matrix transpose.  The boldface means a sequence of elements, e.g., \pmb{$u$} stands for the control sequence.  For a vector $x$, $|x|$ is its absolute value and $||x||$ is its $\mathcal{L}_2$ norm by default. The Euclidean norm of a vector $x \in \mathbb{R}^n$ is denoted by $||x|| = \sqrt{x_1^2 + \cdots + x_n^2}$ whereas $||x||^2_{\Psi}$ , with $\psi$ = $\Psi^{\top} > 0$,  denotes the quadratic form $x^{\top}\Psi x$. $I_m$ is $m$-dimensional unit matrix.

\section{Preliminaries and problem statement}\label{sec:pre}
 We consider the following constrained discrete-time linear time-invariant (LTI) system
\begin{equation}
  \label{eqn:LTI system}
  x(t+1)=Ax(t)+Bu(t), \forall t\in \mathbb{Z}^+
\end{equation}
where $t$ is sampling time, $x \in R^{m}$ is the system states, $u \in R^{n}$ is the control inputs, $A \in R^{{m}\times{m}}$ is the system matrix, and $B \in R^{{n}\times{n}}$ is the input matrix. The pair $(A, B)$ is assumed stabilizable.

The state and control are subject to state and control constraints
\begin{equation}
  \label{eqn:constraints}
  x(t) \in \mathbb{X}, u(t)\in \mathbb{U}, \forall t\in \mathbb{Z}^+.
\end{equation}

Our goal is to compute a sequence of control inputs to steer the system state to origin subject to the constraints. We are interested in formulating a standard model predictive controller (MPC). At sampling time $t$, we solve a constrained finite-time optimal control problem, 
\begin{subequations}\label{eqn:standard MPC}
\begin{align}
\min_{\pmb{u_{0:N-1}}} \quad &\sum_{k=0}^{N-1}{x_k}^{\top}Q{x_k}+ {u_k}^{\top}R{u_k}+{x_N}^{\top}P{x_N}  \\
    \textrm{s.t.} \quad &x_{k+1}=Ax_k+Bu_k,\\
    &x_k\in \mathbb{X}, u_k\in \mathbb{U}, \forall k=0,1,\dots, N-1 \\
    &x_N\in \mathbb{X}_f,\\
    &x_0=x(t)\label{eqn:initial state}
\end{align}
\end{subequations}
where $x(t)$ is the sampled system state at time $t$. Sequence $\pmb{u_{0:N-1}}=\left[\ u_0,u_1,\dots, u_{N-1}\right]^{\top}$ 
is a vector that contains the sequence of control inputs. The symmetric matrices $Q \in R^{{m}\times {m}}$, $R \in R^{{n}\times {n}}$, and $P \in R^{{m}\times {m}}$  define the desired system behavior. The state, terminal, and input constraints are bounded polytopic C-sets $\mathbb{X}$ , $\mathbb{X}_f$ , and $\mathbb{U}$. Without loss of generality, the sets $\mathbb{X}$, $\mathbb{X}_f$, and $\mathbb{U}$ contain their origins as interior points. The set of initial states $x_0$ has a feasible solution depending on the prediction horizon $N$ is called the feasibility region of MPC, denoted by $\mathbb{X}_N$.



Let $\kappa_{MPC}:\mathbb{X}_N\to \mathbb{U}$ denotes the MPC law that is implicitly defined as the solution of the problem (\ref{eqn:standard MPC}). However, the multiparametric solution for MPC is either difficult to obtain or too computationally expensive to deploy on resource-limited hardware. Consequently, we propose to employ a more efficient representation of the control law by leveraging function approximation approaches. In this paper, the deep neural network (DNN) is employed as a parameteric function approximator $\kappa_{NN}(x)$, which can be utilized to approximate the optimal control law $\kappa_{NN}(x)\approx\kappa_{MPC}(x)$. A DNN with $L\in \mathbb{Z}^+(L\geq 2)$ layers represents $\kappa_{NN}(x)$ as a composition of $L$ affine functions $\lambda_i(x):=W_jx+b_i$, each except the last one followed by a nonlinear activation function $h$:
\begin{equation}
  \label{eqn:g function}
  \kappa_{NN}(x) = \lambda_L \circ h \circ \lambda_{L-1} \circ \cdots \circ h \circ \lambda_1(x),
\end{equation}
where $\{W_{1:L},b_{1:L}\}$ are the affine function parameters to be optimized, and $h$ is a nonlinear activation function. In this paper, rectified linear units (ReLUs) are utilized as the activation function, $h(x):=max\{0,x\}$. A DNN with ReLUs is a composition of piecewise affine (PWA) function  on polytopes, and the optimal control law $\kappa_{MPC}$ of the problem (\ref{eqn:standard MPC}) is also a PWA function on polytopes \cite{pascanu2013number}.

Inspired by the command governor\cite{bemporad1997nonlinear}, we propose an approach named safety governor to guarantee the safety of the approximated controller, denoted by $SG$: \begin{equation}
  \label{eqn:nonlinear control}
  \kappa_{SG}(x)=SG(\kappa_{NN}(x))
\end{equation} 

In summary, the objective is to design, via DNN and safety governor, a nonlinear state feedback $u(t)=\kappa_{SG}(x(t)), t\in \mathbb{Z}^+$ steers the states of system (\ref{eqn:LTI system}) to the origin such to the constraints (\ref{eqn:constraints}). A further objective is that the nonlinear feedback (\ref{eqn:nonlinear control}) reduce to a linear well-tuned feedback whenever this is compatible with the constraints.

\section{Main result}
In this section, we propose a learning-based explicit MPC,  safety governor, and validation of the proposed controller. 
\subsection{Learning-based explicit MPC}
In MPC problem (\ref{eqn:standard MPC}),  terminal cost ${x_N}^{\top}Px_N$ and terminal constraints $\mathbb{X}_f$ are introduced to simplify constrained infinite horizon LQR problem by restricting the optimization to a finite horizon $N$. For computation reasons, an achievable goal is to steer the initial state $x_0$ to a neighborhood of the origin $\Sigma$. 
Consider a  stabilizing LTI  state feedback to meet the desire for linear control when $x(t)\in\Sigma$:
\begin{align}\label{eqn:LTI  state feedback}
u(t)=Kx(t), \forall t\in \mathbb{Z}^+
\end{align}
designed so as to provide a satisfactory, or optimal in some sense (e.g. LQ), control performance to  system (\ref{eqn:LTI system}) in the absence of constraints. The corresponding asymptotically stable nominal closed-loop system is denoted as
\begin{align}\label{eqn:nominal closed-loop}
x(t+1)=(A+BK) x(t), \forall t\in \mathbb{Z}^+.
\end{align}
Without loss of generality, $A + BK$ is assumed 
has all eigenvalues strictly inside the unit circle.

To characterize the target set $\Sigma$, it have to introduce the notion of maximal constraint admissible set \cite{gilbert1991linear}. 
\begin{definition}\label{def:positively invariant}
A set $\Sigma \in \mathbb{R}^{m} \subset \mathbb{X} $ is called  constraint admissible set for system (\ref{eqn:nominal closed-loop}), if and only if  $\forall x\in \Sigma$, $Kx\in \mathbb{U}$ and $(A+BK) x\in \Sigma$.
\end{definition}
Then, the following proposition holds.
\begin{proposition}\label{prop:max positively invariant}
The maximal constraint admissible set $\Sigma_\infty$ is constraint admissible, and contains all constraint admissible sets. In other words, the maximal constraint admissible set is the union of all constraint admissible sets.
\end{proposition}

The maximal constraint admissible set can be determined as the limit of a recursive process \cite{gilbert1991linear}:
\begin{align}\label{eqn:recursive process}
\Sigma_{i+1}=Pre(\Sigma_i)\cap \Sigma_i, \Sigma_0=\mathbb{X}
\end{align}
where $Pre(\Sigma)=\{x | \exists u=Kx\in \mathbb{U},s.t.\ Ax+Bu\in\Sigma  \}$ denotes the predecessor set. If and only if 
 $\Sigma_{i^*+1}=\Sigma_i^*$ for some $i^*\in \mathbb{Z}^+$, then the maximal constraint admissible set $\Sigma_{\infty}=\Sigma_{i^*}$. 

The LQ state feedback law is adopted in this paper: 
\begin{align}\label{eqn:compute K}
K = -(R + B^{\top} PB)^{-1}(B^{\top}PA)  
\end{align}
where the matrix $P$ is the solution to the discrete Algebraic Riccati Equation \cite{1995Dynamic}:
\begin{align}\label{eqn:compute P}
P = A^{\top} P A + Q - A^{\top} P B (B^{\top} P B + R)^{-1} B^{\top}PA  
\end{align}
which corresponds to the optimal cost-to-go after $N$ steps for the unconstrained infinite-horizon LQR problem. 

Clearly, the definition of  the maximal constraint admissible set means that when the system states are steered to the neighborhood of the origin, the nonlinear feedback will  reduce to the linear
feedback naturally, without any constraints violation. In other words, 
\begin{align}\label{eqn:mpc xf}
\kappa_{MPC}(x)=Kx, \forall x\in \Sigma_\infty.
\end{align}

\begin{figure}[!htb]
  \centering
  \includegraphics[width=0.95\linewidth]{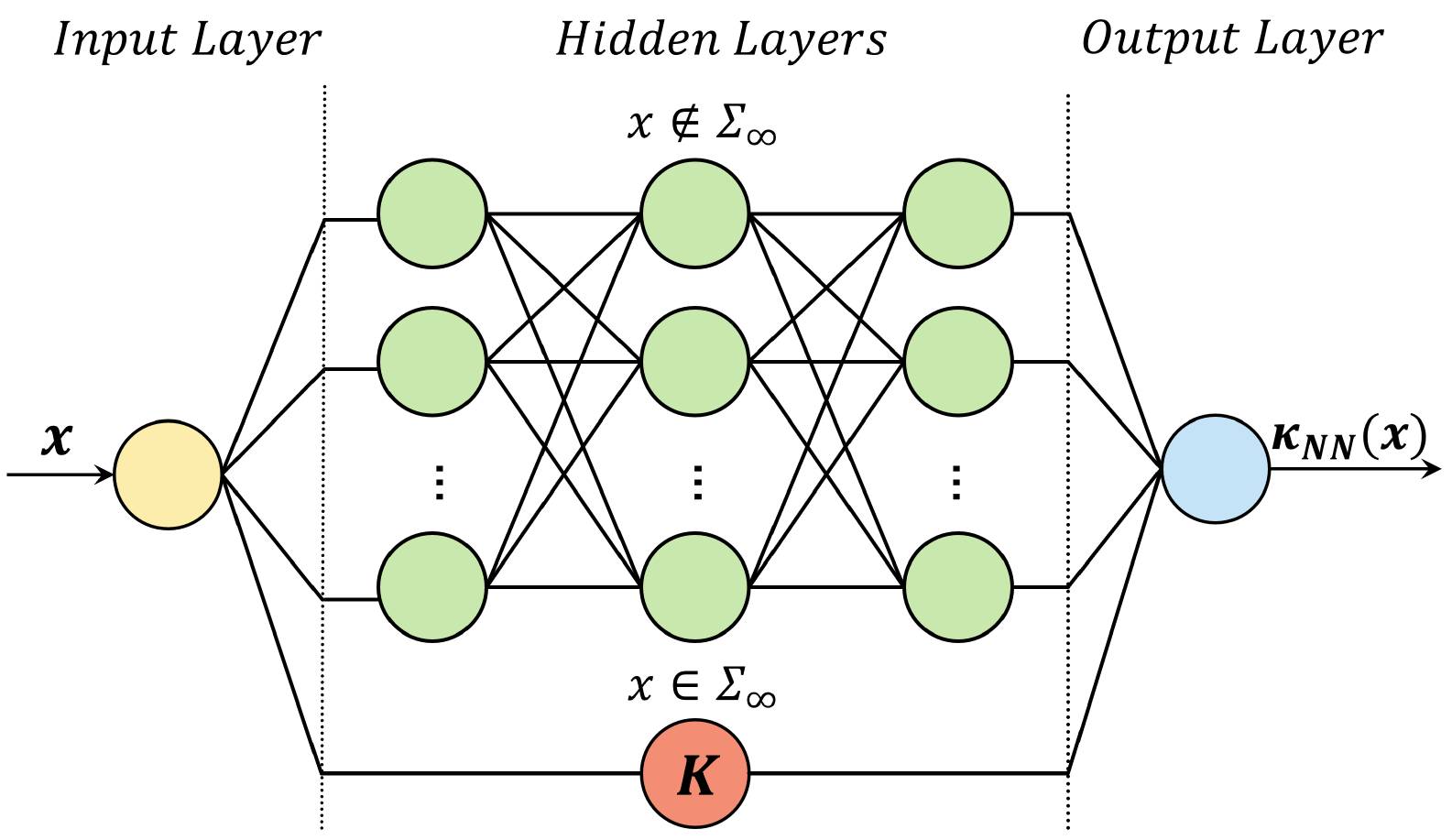}
  \caption{Diagram of the learning-based explicit MPC structure}
  \label{fig: NN structure proposed}
\end{figure}

On the basis of this consideration, a novel learning-based explicit MPC structure is proposed, where the learning-based explicit MPC is  reformulated as a dual-mode scheme: 
\begin{align}\label{eqn:nn structure}
\hspace{0.2cm} \kappa_{NN}(x) & \\ 
& =\begin{cases}
Kx,&x\in \Sigma_\infty; \\
\lambda_L \circ h \circ \lambda_{L-1} \circ \cdots \circ h \circ \lambda_1(x),&x\notin \Sigma_\infty. 
\end{cases}\nonumber
\end{align}
Figure \ref{fig: NN structure proposed} shows the proposed learning-based explicit MPC structure.

The proposed learning-based explicit MPC structure (\ref{eqn:nn structure}) has ability to erase the approximation errors while the system states had been steered into the neighborhood of the origin. The numerical experiment in next section will prove this more directly. Then, we adopt a supervised learning  method to train the learning-based explicit MPC. First, we generate an arbitrary number $n\in \mathbb{Z}^+$ of samples  $(x^{(i)},\kappa_{MPC}(x^{(i)})) \in (\mathbb{X}_N\times \mathbb{U})$, where $i=1,\dots,n$. We then use the mean square error (MSE) between the control inputs of  MPC and learning-based explicit MPC as the loss function:
 \begin{align}\label{eqn:loss function}
Loss=\frac{1}{n}\sum_{i=1}^n(\kappa_{MPC}(x^{(i)})-\kappa_{NN}(x^{(i)}))^2. 
\end{align}

\subsection{Safety governor}

A trained learning-based explicit MPC is difficult to guarantee its safety due to the existence of approximation errors. In this paper, inspired by the command governor (CG) \cite{bemporad1997nonlinear}, a safety governor is constructed to ensure that learning-based explicit MPC satisfies all the state and input constraints.

Consider the equilibrium state $x_s$ and its corresponding control input $u_s$ of system (\ref{eqn:LTI system}):
 \begin{align}\label{eqn:equilibrium states}
x_s=Ax_s+Bu_s.
\end{align}
The equation can be reformulated to
 \begin{align}\label{eqn:zero space}
\begin{pmatrix} A - I & B \end{pmatrix} \begin{pmatrix} x_s \\ u_s \end{pmatrix} = 0.
\end{align}
Then the equilibrium state and its control input can be parameterized as
 \begin{align}\label{eqn:zero space parameterization}
\begin{pmatrix} x_s \\ u_s \end{pmatrix}=\begin{pmatrix} M_x \\ M_s \end{pmatrix} \gamma
\end{align}
where $x_s=M_x\gamma$ and $u_s=M_u\gamma$. Then,
 \begin{align}\label{eqn:gamma set}
\Gamma\triangleq\{\gamma:M_x\gamma\in X\ \text{and}\ M_u\gamma\in U\}
\end{align}
is the set of statically admissible commands, i.e. the set of $\gamma$ which satisfy the constraints on $x$ and $u$ in steady state. For each $\gamma \in \Gamma$,
 \begin{align}\label{eqn:cons feasible set aug}
\Sigma_{\infty}(\gamma)\triangleq\{\nonumber&x_0:x_{k+1}=Ax_k+B(Kx_k+M_\gamma\gamma)\in \mathbb{X},   \\ 
& Kx_k+M_\gamma\gamma\in \mathbb{U}, \forall k\in \mathbb{Z}^+\}
\end{align}
is the maximal constraint admissible set under the control law $u=K(x-x_s)+u_s=Kx_k+(M_u-KM_x)\gamma=Kx_k+M_\gamma\gamma$, where $M_\gamma=M_u-KM_x$. Similar to the recursive procedure in (\ref{eqn:recursive process}), $\Sigma_{\infty}(\gamma)$ can be finitely determined \cite{gilbert1991linear}. Clearly, 
\begin{align}\label{eqn:Sigma equation}
\Sigma_\infty=\Sigma_{\infty}(0).
\end{align}
Let 
\begin{align}\label{eqn:Sigma union}
\Sigma_\infty(\Gamma)\triangleq \bigcup_{\gamma \in \Gamma}\Sigma_{\infty}(\gamma),
\end{align}
which is an augmented maximal constraint
feasible set, and also a positive invariant set \cite{gilbert1991linear}.

For any given $x(t) \in \Sigma_\infty(\Gamma)$ and control input $\kappa_{NN}(x(t))$ of learning-based explicit MPC. Then, the safety governor $SG$ determine the control inputs $u(t)$ by solving the following problem:
\begin{subequations}\label{eqn:safety governor}
\begin{align}
\min_{u(t)} \quad &||u(t)-\kappa_{NN}(x(t))||_s  \\
    \textrm{s.t.} \quad &u(t)=Kx(t)+M_\gamma\gamma(t),\label{eqn:final control}\\
    &x(t)\in \Sigma_{\infty}(\gamma(t)),\gamma(t)\in\Gamma
\end{align}
\end{subequations}
where $s > 0$. Then, we can formalize the following theorem via the above safety governor. \begin{theorem}\label{theorem}
 Consider the system (\ref{eqn:LTI system}) with constraints (\ref{eqn:constraints}), control input is defined by law (\ref{eqn:nonlinear control}) and $x(0) \in \Sigma_{\infty}(\Gamma) \subset{\mathbb{X}}$. Then $x(t) \in \mathbb{X}$ and $u(t)\in\mathbb{U} $ are defined for all $t\in \mathbb{Z}^+$. 
\end{theorem}

\noindent\textit{Proof}:
When $t=0$, the problem (\ref{eqn:safety governor}) is naturally feasible for $x(0)\in \Sigma_{\infty}(\Gamma)\subset\mathbb{X}$ from (\ref{eqn:cons feasible set aug}) and (\ref{eqn:Sigma union}). In other words, there exists $u(0)=Kx(0)+M_\gamma\gamma(0)\in\mathbb{U}$ with $\gamma(0)\in\Gamma$. When $t=k$, the problem (\ref{eqn:safety governor}) is assumed feasible for $x(t)\in\Sigma_{\infty}(\gamma(t))\subset\Sigma_{\infty}(\Gamma)\subset\mathbb{X}$. There exists $u(t)=Kx(t)+M_\gamma\gamma(t)\in\mathbb{U}$ with $\gamma(t)\in\Gamma$. When $t=k+1$, $x(t+1)=Ax(t)+B(Kx(t)+M_{\gamma}\gamma(t))$. Since $x(t)\in\Sigma_{\infty}(\gamma(t))\subset\Sigma_{\infty}(\Gamma)\subset\mathbb{X}$,
$x(t+1)\in\Sigma_{\infty}(\gamma(t))$. The problem (\ref{eqn:safety governor}) is feasible for $x(t+1)\in\Sigma_{\infty}(\gamma(t))\subset\mathbb{X}$. There exists $u(t+1)=Kx(t+1)+M_\gamma\gamma(t+1)\in\mathbb{U}$ with $\gamma(t+1)=\gamma(t)\in\Gamma$. In conclusion, Theorem \ref{theorem} is proved.  $\hfill\blacksquare$



Given the proposed learning-based explicit MPC and safety governor, the overall procedure for learning MPC law is summarized in Algorithm \ref{algo:whole procedure}.
\begin{algorithm}[h]
		\caption{Learn MPC law}
		\begin{algorithmic}[1]
			\renewcommand{\algorithmicrequire}{\textbf{Input:}}
			\renewcommand{\algorithmicensure}{\textbf{Output:}}
			\REQUIRE System (\ref{eqn:LTI system}), matrices $Q,R$, and horizon $N$
			\ENSURE  Nonlinear state
                feedback $u=SG(\kappa_{NN}(x))$ (\ref{eqn:nonlinear control})\\
			\textit{Initialization}: 
            Construct network (\ref{eqn:nn structure}) with $K$ obtained from (\ref{eqn:compute K}), where $P$ is computed by solving (\ref{eqn:compute P});
			\STATE  Sample the MPC problem (\ref{eqn:standard MPC}) to generate an arbitrary number $n \in\mathbb{Z}^+$  as train set: $(x^{(i)},\kappa_{MPC}(x^{(i)})) \in (\mathbb{X}_N\times \mathbb{U})$, where $i=1,\dots,n$;
            
			\STATE Learn $\kappa_{NN}(x)\approx\kappa_{MPC}(x)$, with network (\ref{eqn:nn structure}) and loss function (\ref{eqn:loss function});

            \STATE Obtain the safety governor $SG$ from (\ref{eqn:cons feasible set aug}) and (\ref{eqn:Sigma union});

            \STATE Validate the closed-loop control performance of law $u=\kappa_{SG}(x)$. 

		\end{algorithmic}
		\label{algo:whole procedure}
	\end{algorithm}

\begin{remark}
    The feasible region of Problem (\ref{eqn:standard MPC}) is not the same as that of Problem  (\ref{eqn:safety governor}) \cite{gilbert2011constrained}, i.e., $\mathbb{X}_N\neq \Sigma_{\infty}(\Gamma)$. For the proposed Theorem \ref{theorem}, $x_0 \in \Sigma_{\infty}(\Gamma) \subset{\mathbb{X}}$.
\end{remark}

\section{Numerical experiment}
\noindent \textbf{Example 1.} Consider a double integrator system:
\begin{align}
    x(t+1)= \begin{bmatrix} 1 & 0.5 \\ -0.1 & 0.9 \end{bmatrix} x(t) + \begin{bmatrix} 1 \\ 0 \end{bmatrix} u(t), t\in \mathbb{Z}^+
\end{align}
where $ \mathbb{U} = \{u \in \mathbb{R} :  u\in [-1,1] \}$, $\mathbb{X} = \{x \in \mathbb{R}^2: x \in [-5, 5] \times [-5, 5]\}$, and  $\mathbb{X}_f:=\Sigma_\infty$.
We consider the MPC law $\kappa_{MPC}$ to be the solution to the problem (\ref{eqn:standard MPC}) with  $Q = I_2$, $R = I_1$, $P = \begin{bmatrix} 1.71 & -0.26 \\ -0.26 & 5.53 \end{bmatrix}$, $K=\begin{bmatrix} -0.64 & -0.23  \end{bmatrix}$, and predictive horizon $N = 10$. 

\textit{Step 1 (Data sampling):} $1\times10^2$ feasible samples of data pairs $(x, \kappa_{MPC}(x))$ are generated as training set, where $x\in \mathbb{X}_N$ and $\kappa_{MPC}(x)$ is obtained by solving the MPC problem (\ref{eqn:standard MPC}), using CVXPY 1.5.2 with Python 2.8.0.

\textit{Step 2 (Learning-based explicit MPC):} 
A fully connected feedforward NN with 2 neurons in input layer and three further hidden layers with 20 neurons each is utilized to approximate the MPC law. In the output layer, linear activations are used with one neuron. The learning-based explicit MPC was trained under loss function (\ref{eqn:loss function}) by using the deep learning toolbox PyTorch on an NVIDIA 3090 GPU. The Adam optimizer  is used to optimize our model. The initial learning rate is set to 0.001 and remains unchanged throughout the training phase that goes through 1000 epochs.

\begin{figure}[!htb]
  \centering
  \includegraphics[width=0.95\linewidth]{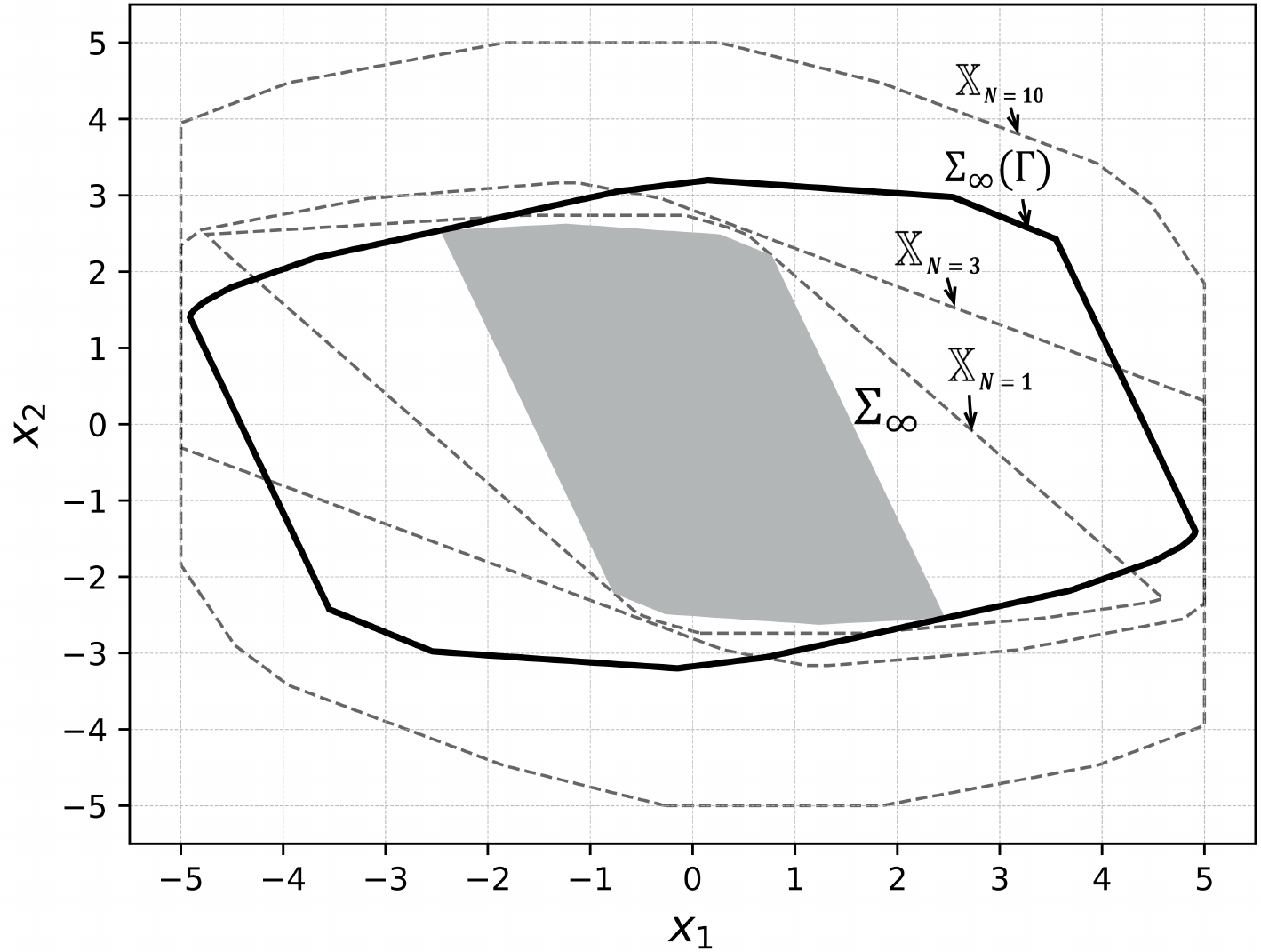}
  \caption{ Comparison of feasibility regions for different approaches}
  \label{fig: feasibility region comparison}
\end{figure}

\textit{Step 3 (Safety governor):} 
We obtain the safety governor $SG$ from (\ref{eqn:cons feasible set aug}) and (\ref{eqn:Sigma union}). For different control approaches, it is interesting to compare their feasible regions. Figure  \ref{fig: feasibility region comparison} shows the feasible regions of our approach, MPC with $N=1$, $N=3$ and $ N=10$: $\Sigma_{\infty}(\Gamma)$, $\mathbb{X}_{N=1}$, $\mathbb{X}_{N=3}$, and $\mathbb{X}_{N=10}$. Notice that our approach has a size of the feasible region close to that of the MPC with $N=3$, but with a completely different shape. The MPC with $N=10$ has a significantly larger feasible region, but this corresponds to an increase in its online computation time, which will be discussed in more detail later.

\begin{figure}[!htb]
  \centering
  \includegraphics[width=0.95\linewidth]{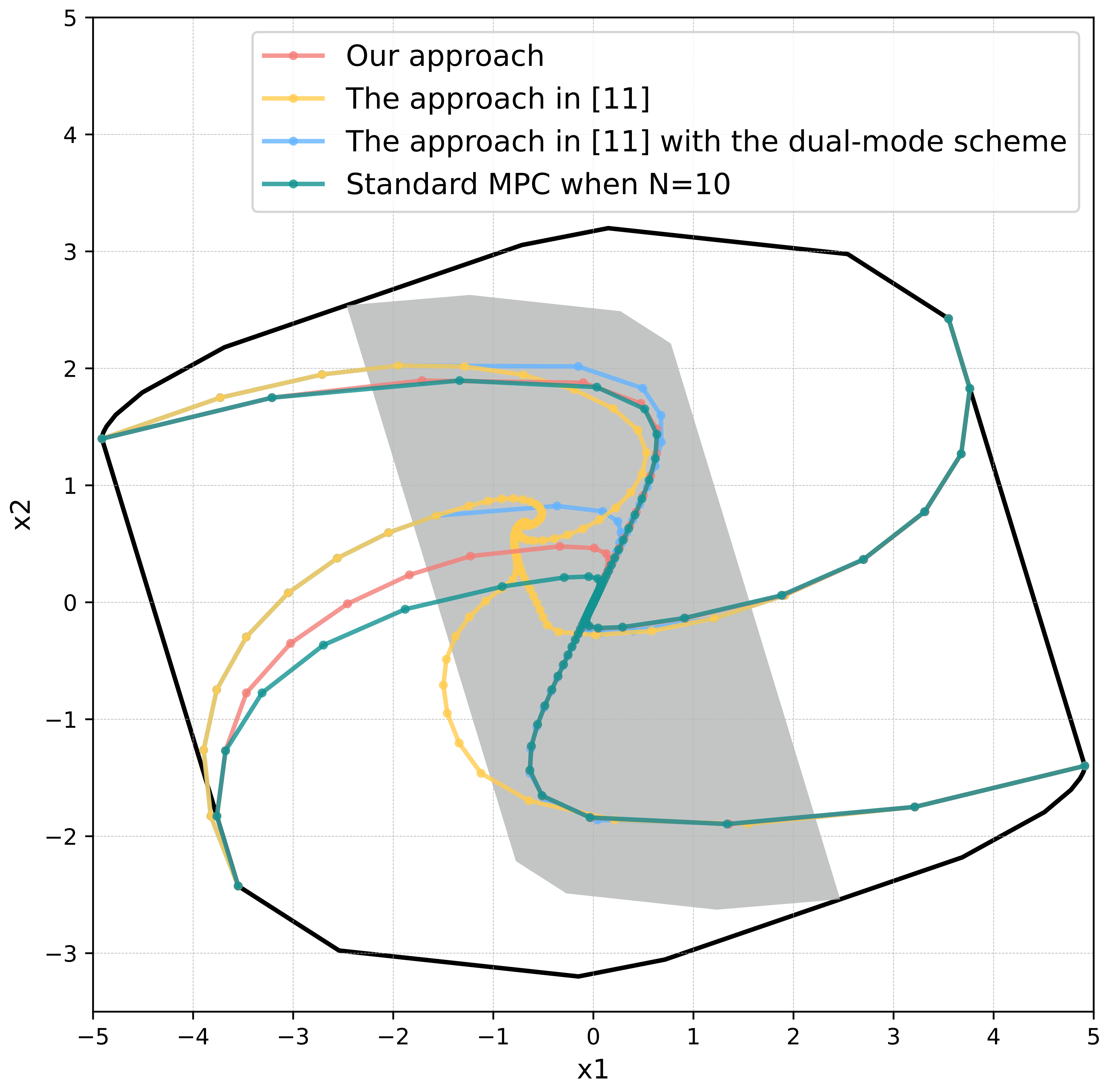}
  \caption{Comparison of system state trajectories for different approaches}
  \label{fig: trajectories comparison}
\end{figure}

\textit{Step 4 (Validation):} 
This paper contrasts \cite{chen2018approximating} projecting control input of the DNN onto the feasible region of MPC: $u=\text{arg} \min_{u} \||u-\kappa_{NN}(x)||_s$, such that $u\in\mathbb{U},x^+=Ax+Bu\in \mathbb{X}_N,\forall x\in \mathbb{X}_N$. Figure \ref{fig: trajectories comparison} shows the system state trajectories derived by applying different control approaches with several feasible region vertices as initial states. The red trajectories represent our approach, the yellow trajectories represent the approach in \cite{chen2018approximating}, the blue trajectories represent the approach in \cite{chen2018approximating} with dual-mode scheme, and the green trajectories represent the MPC with $N = 10$. Our approach ensures that the system states remain within the safe region while converging closer to the origin,  owing to the safety governor proposed. In addition, the existence of the approximation error does not affect the system state to be steered to the origin owing to the dual-mode scheme proposed. 

Table \ref{table: Computation time for different approaches} compares the computational time required for different approaches across 50 steps. This comparison verifies the superiority of our approach in terms of computational time. The computational time of standard MPC with $N=10$ is also much longer than the other approaches.
\begin{table}[h]
\caption{Comparison of computational time for different approaches}
\label{table: Computation time for different approaches}
\begin{center}
\setlength{\tabcolsep}{15pt} 
\renewcommand{\arraystretch}{1.4} 
\begin{tabular}{|c||c|}
\hline
 & Computational time \\
\hline 
Standard MPC with N=1 & 1.21 [s] \\
\hline
Standard MPC with N=3 & 3.25 [s] \\
\hline
Standard MPC with N=10 & 8.53 [s] \\
\hline
The approach in \cite{chen2018approximating} & 1.21 [s] \\
\hline
Our approach & 1.20 [s] \\
\hline
\end{tabular}
\end{center}
\end{table}

\noindent \textbf{Example 2.} Consider the following 4-dimensional system:
\begin{align}
    A= \begin{bmatrix} 0.7 & -0.1 &0&0 \\ 0.2 & -0.5&0.1&0\\0 &0.1&0.1&0\\0.5&0&0.5&0.5 \end{bmatrix}, B=\begin{bmatrix} 0.1 \\ 0.1\\0.1\\0.1 \end{bmatrix}
\end{align}
with constraints: $ \mathbb{U} = \{u \in \mathbb{R} :  u\in [-2,2] \}$, $\mathbb{X} = \{x \in \mathbb{R}^4: x \in [-5, 5] \times [-5, 5]\times [-1, 1]\times [-1, 1]\}$. Parameters $Q=I_4$, $R=I_1$ and $N=10$. 

The simulation follows the same procedure as the double integrator system, with only minor modifications in the following implementation details: (1) A larger dataset containing $5\times10^2$  feasible sample pairs $(x, \kappa_{MPC}(x))$ is generated; (2) The input layer dimension is increased to four neurons; (3) The neural network architecture is enhanced with six hidden layers.

\begin{figure}[!htb]
  \centering
  \includegraphics[width=0.95\linewidth]{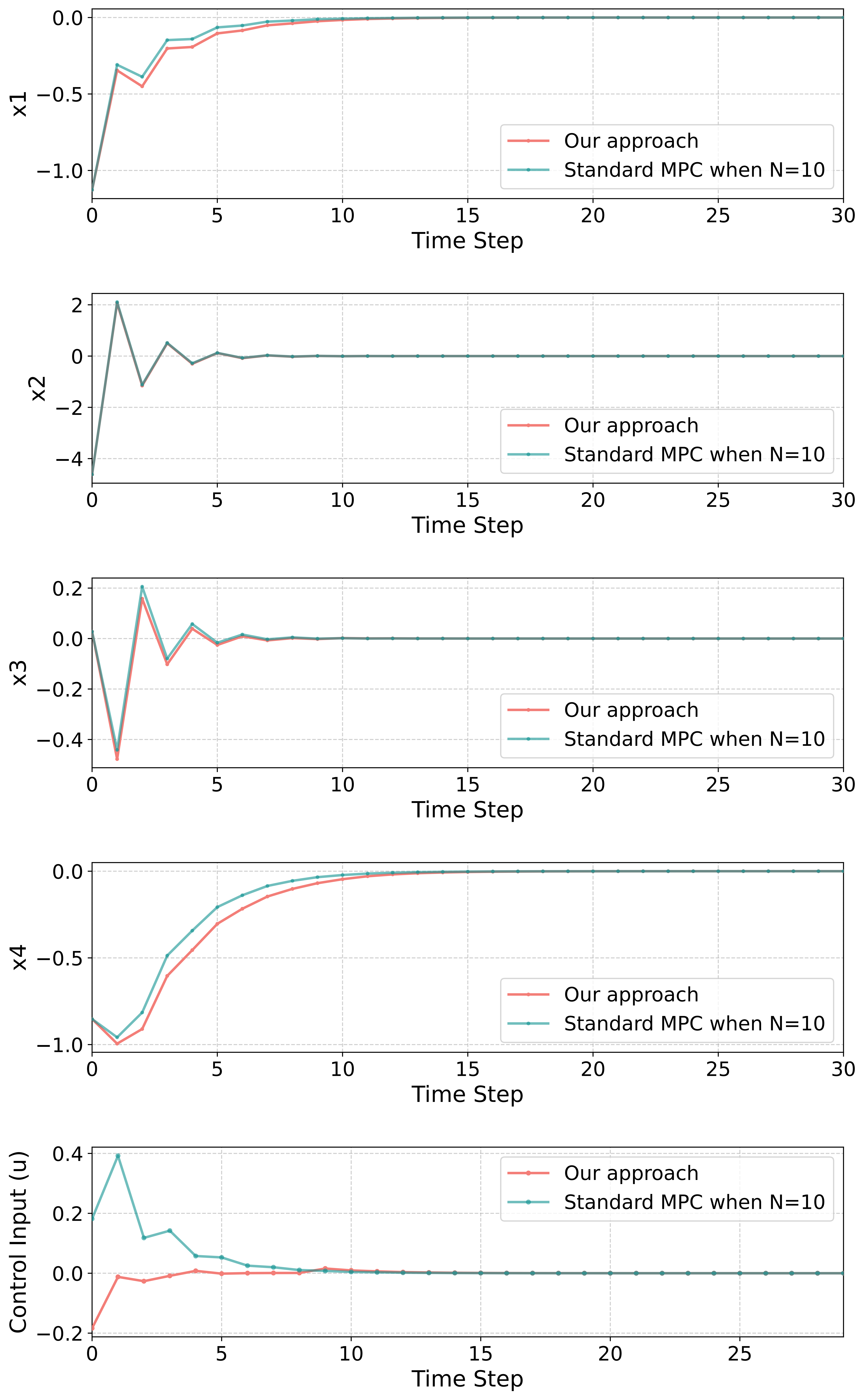}
  \caption{Comparison of state components and control inputs for different approaches}
  \label{fig: System states and control inputs comparison}
\end{figure}

\textit{Simulation results:} In high-dimensional system, the determination of the feasible region $\mathbb{X}_N$ for MPC is an NP-complete problem \cite{2008On}. Although there exists approach to approximate, which may loss the feasibility of the approach in \cite{chen2018approximating}. The determination of feasible region $\Sigma_\infty(\Gamma)$ is much easier. Figure \ref{fig: System states and control inputs comparison} shows the control inputs and the corresponding system state components under our approach and standard MPC with $N=10$ while $x_0=\begin{bmatrix}-1.12& -4.62&0.03&-0.85\end{bmatrix}^{\top}$.  Our approach shows rewarding control performance in this high-dimensional system, constraints are not violated, and is able to steer the system states to the origin even with approximation errors.

\section{Conclusion}
We have proposed a learning explicit MPC  via deep neural network for linear time-invariant systems. Our approach can reduce online computation while guarantee the rewarding control performance and safety. Leveraging the maximal constrained feasible set, we propose a learning-based explicit MPC structure to erase the approximation errors nearby the origin of system. Following the parameterization of system states and control inputs using equilibrium states, a safety governor is constructed to ensure that learning-based explicit MPC satisfies all the state and input constraints. A numerical experiment is proposed to verify the rewarding control performance of our approach.

\bibliographystyle{unsrt}
\bibliography{arxiv}

\end{document}